# Phonon-mediated closing of topological Floquet gaps in graphene: Non-phenomenological analysis

Royi Ledermann[1,2,*], Rave Hanoch Saadon[1,2,*], Adam Herling[1], and Ofer Neufeld[1,†]

[1]Technion Israel Institute of Technology Faculty of Chemistry, Haifa 3200003, Israel.

[2]Technion Israel Institute of Technology, Faculty of Electrical and Computer Engineering, Haifa 3200003, Israel.

[*]These authors contributed equally to this work.

[†]Corresponding author E-mail: ofern@technion.ac.il

Floquet band engineering has been intensively studied for its potential to control material properties via laser driving. In particular, Floquet topological states have been measured on the surface of $Bi_2Se_3$. Nonetheless, the original prediction of Floquet topological bands in graphene remains unobserved, with works only measuring non-topological Floquet-Bloch states or indirect features. Here, we theoretically explore graphene irradiated by circular lasers with non-phenomenological electron-phonon (e-ph) coupling. We show that e-ph interactions substantially broaden Floquet bands due to graphene's large zero-point motion, as expected from phenomenological treatments. However, we further find that larger displacement Γ optical phonons reduce Floquet gaps by about half, even in absence of instrumentation broadening, and can also induce additional features like flat bands. These phonons 'enter the gap' and blur photoemission signatures. Including non-Γ phonons closes the gap and greatly reduces visibility. When these modes couple to reasonably expected instrumentation broadening, the effect is significantly exacerbated. Our results shed light on the missing Floquet topological gaps and also lead to possible mitigation strategies: (i) Pre-pumping coherent phonons to counteract blurring, or (ii), transition to a Dirac system with more favorable phonon statistics. Moreover, our real-space analysis proposes that Floquet topological physics is alive in localized regions, such that properties of interest like transport should be accessible. We expect this work to impact experimental analysis and lead to set-ups where graphene Floquet topology might finally be directly observed.

Over the past decade laser driving has emerged as a key paradigm in material control and design. The basic idea of this approach is driving quantum materials out-of-equilibrium (in either electronic or phononic landscapes) to induce phase transitions [1,2] and cause the system to enter light-dressed states (Floquet) [3–5]. This gives rise to various tunable control knobs such as the laser frequency, polarization, carrier components [6], etc. For example, implementations include light-induced superconductivity [7,8], optically-controlled magnetism [9–15], Floquet band engineering [16–21], and topological anomalous Hall physics [22–25].

Within this field, a key work is that by Oka and Aoki for Floquet topological gap opening in graphene dressed by circularly-polarized light [26]. Here the circular laser breaks time-reversal symmetry, removing the Kramers degeneracy at the Dirac points causing gap opening, while also leading to nonzero Chern numbers in the associated Floquet quasi-energy bands (topological bands). Since its prediction, great effort was invested to observe the gap opening, but attempts have generally failed. Indeed, gap opening was measured in $Bi_2Se_3$ surface states that also have Dirac dispersion [27], but in monolayer graphene only non-topological Floquet bands and gaps were observed [28–30]. Several works attempted to understand this inconsistency and identify potentially improved conditions for observation [31–37], or alternative schemes of observation [38,39]. Largely, carrier scattering is believed to be a main limitation [37], and *ab-initio* calculations have shown both electron-electron and electron-phonon (e-ph) interactions play a role [36]. However, an analysis of the gap visibility for varying experimentally relevant laser regimes has not been carried out, and it's not clear which scattering mechanism is most relevant and why. In this context, it's noteworthy that indirect signatures of the topological bands have not had such issues, and both anomalous Hall [22,40] and anomalous optical Hall [24] physics have been observed, as well as photocurrent circular dichroism [24], as expected from Chern insulators.

Here, we theoretically study Floquet topological phases in graphene with non-phenomenological electron-phonon (e-ph) interactions. This goes beyond most past works where only phenomenological dephasing was considered, or an e-ph microscopic theory [29] that does not include phonon spatial/momentum dependence (apart from the extensive analysis of Schüler et al. [36]). These treatments limit impact of e-ph scattering to an expected spectral broadening. By incorporating thermally-occupied phonons in the static limit with spatial and momentum dependence, we show that over-simplified treatments fail to capture interactions that blur the gap, reduce its size, and can ultimately close it. We uncover the physical mechanism behind this effect: large lattice displacement Γ modes (with ~10% occurrence) activate a mechanism that allows them to 'enter the gap' and reduce its size by half. Non-Γ phonons further enhance this effect, greatly reducing gap visibility. When these phonon modes couple to instrumentation broadening gap visibility can be fully blurred. Beyond proposing a novel explanation for the lack of Floquet gap observability in graphene, our results have striking implications. First, they imply Floquet topology is live and well as long as phononic averaging is small (in local regions), and that the states should be observable with local probes [39]. Second, our work proposes natural mitigation strategies for measurements, either pre-pumping a coherent phonon mode [41–48], choosing a material with improved phonon statistics, or using higher-power lasers.

Let us begin by describing our methodology. Graphene electrons are described within a second nearest neighbor (NN) tight-binding (TB) Hamiltonian. This captures correct Dirac cone physics, including effects such as broken particle hole-symmetry. The Hamiltonian for the equilibrium lattice reads:

$$H_0(\mathbf{k}) = \begin{pmatrix} t_2 f_2(\mathbf{k}) & t_1 f_1(\mathbf{k}) \\ t_1 f_1^*(\mathbf{k}) & t_2 f_2(\mathbf{k}) \end{pmatrix} \tag{1}$$

where $t_1$ and $t_2$ are the first and second NN hopping amplitudes, and $f_{1/2}(\mathbf{k})$ are respective structure functions: $f_1(k) = \sum \exp(-i\boldsymbol{\delta}_{i,1} \cdot \mathbf{k})$, and $f_2(k) = \sum \exp(-i\boldsymbol{\delta}_{i,2} \cdot \mathbf{k})$, with $\boldsymbol{\delta}_{i,m}$ the NN vectors of $m$'th order, and the sum runs over all connections. A laser dressing field within the electric-dipole approximation drives the system into a Floquet state, described via Peierls substitution: $\mathbf{k} \to \mathbf{k}(t) = \mathbf{k} + \mathbf{A}(t)$, where $\mathbf{A}(t) = \frac{E_0}{\omega}\cos(\omega t)\,\hat{\boldsymbol{e}}$, with $E_0$ the electric field amplitude (connected to the peak power, $I_0$), $\omega$ the carrier frequency, $\hat{\boldsymbol{e}}$ a circularly-polarized unit vector, and we employ atomic units throughout. The time-dependent Hamiltonian is then $H(\mathbf{k}, t) = H_0(\mathbf{k}(t))$. Floquet quasi-energy bands are calculated by expanding the Floquet Hamiltonian ($\mathcal{H}_{\mathcal{F}} = -i\partial_t + H(\mathbf{k}, t)$) in the extended space of time-periodic functions [49] (see refs. [50–52] for more details). $\mathcal{H}_{\mathcal{F}}$ is numerically diagonalized while convergence is confirmed (see SI for all details).

Up to this point, the theoretical approach is standard. To describe e-ph interactions non-phenomenologically, we follow our recent implementation [53]. We thermally occupy optical phonons at Γ by inducing lattice distortions that uphold Bose-Einstein statistics (see details in SI). The Hamiltonian is then a function of the phonon displacement vector, $\Delta\mathbf{R}$:

$$H_0(\mathbf{k}, \Delta\mathbf{R}) = \begin{pmatrix} t_2 f_2(\mathbf{k}) & \tilde{f}_1(\mathbf{k}, \Delta\mathbf{R}) \\ \tilde{f}_1^*(\mathbf{k}, \Delta\mathbf{R}) & t_2 f_2(\mathbf{k}) \end{pmatrix} \tag{2}$$

with $\tilde{f}_1$ a generalized structure function that includes exponentially-modulated hoppings, $\tilde{t}_1(\Delta\mathbf{R})$, that depend on the distorted NN vectors, $\tilde{\boldsymbol{\delta}}_{i,1}$ (see SI). Thus, each fixed $\Delta\mathbf{R}$ in this static phonon approximation defines an electronic Hamiltonian, for which the approach above is employed. We sample the phonon distribution by ensemble averaging resulting Floquet bands from each $\Delta\mathbf{R}$. This assumes ARPES bands build up incoherently from different crystal regions that locally exhibit distortions. Each $\Delta\mathbf{R}$ simulation is denoted as a 'snapshot', and the simulation is converged with snapshot number. In order to construct the full spectrogram, each Floquet band from each snapshot is replaced by a Gaussian function that decays in the energy axis with a typical width $\sigma_E$ (broadening associated with the ARPES probe duration), and all snapshots and bands are summed to obtain a 2D image of a spectral function. Surprisingly, large number of snapshots is required to obtain converged spectra (typically >5000), which differs from cases like HHG [53] (see SI for details).

Having outlined the key methodology, let us turn to the analysis. Employing this scheme, Fig. 1 presents Floquet bands simulated in two typical experimental conditions recently employed, and where Floquet gaps are expected with a size of 0.091 eV ($I_0$=1.2×10$^{10}$ W/cm$^2$, λ=1907.5nm) [29], and 0.0757 eV ($I_0$=4.3×10$^9$ W/cm$^2$, λ=2530nm) [30]. Note that these conditions greatly differ from those in ref. [36] where much more intense and high-frequency (800nm) pump was used. For start, we consider negligible instrumentation broadening (ideal conditions, $\sigma_E$= 0.01 eV, corresponding to probe durations of ~400 fs). Figures 1(a,b) show the resulting energy diagrams along $k_x$ dispersion, while $k_y$ plots are delegated to the SI and show similar features. Note that the color scale does not directly connect to ARPES; instead, it represents a probability connected to phonon occupations, on top of which additional electronic occupation and matrix elements should be added. In practice, electrons occupy mostly states close to the original Dirac cone, which is why we focus our analysis on that region.

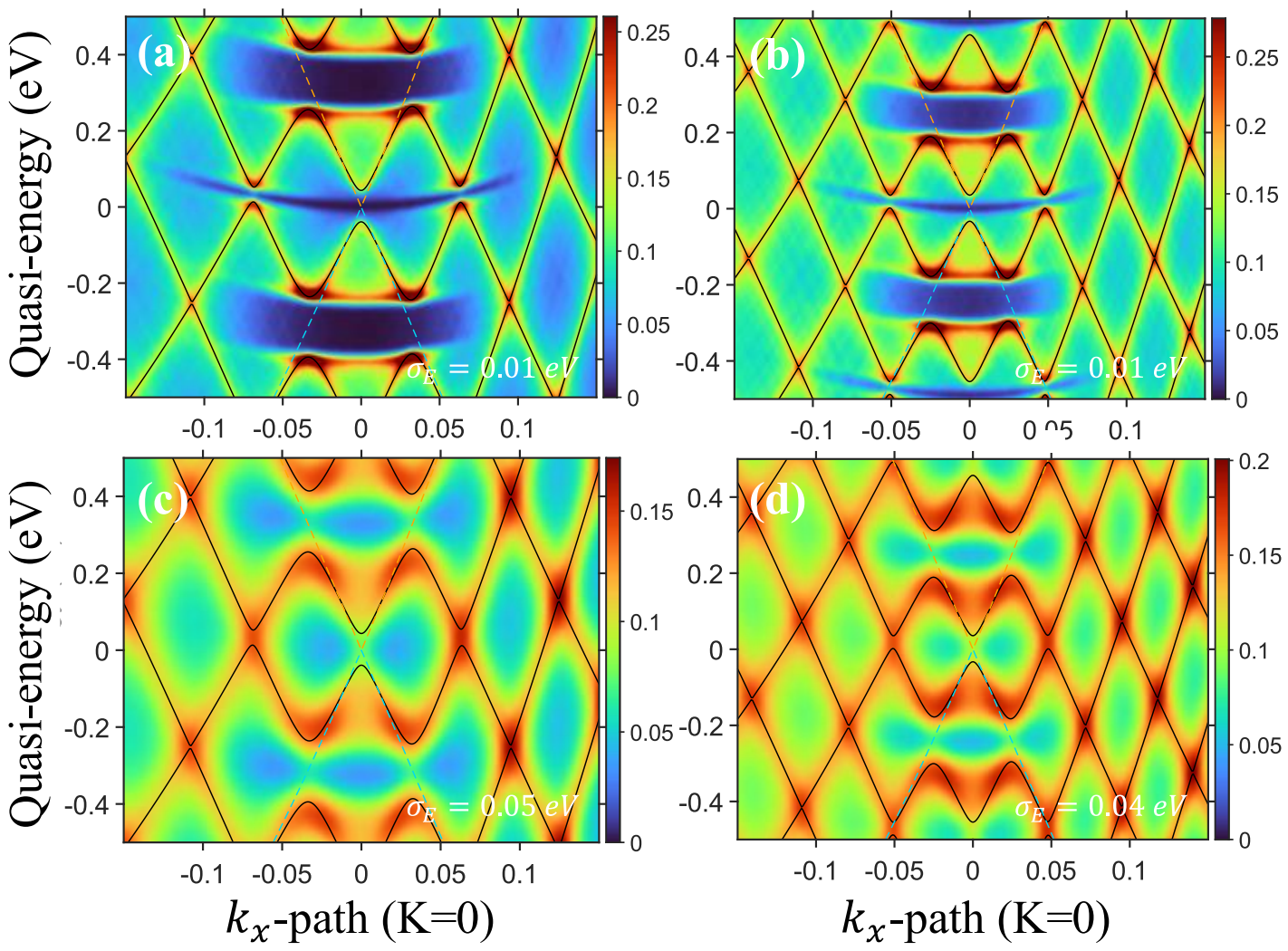

**Fig. 1.** (a) Floquet quasi-energy spectral function including Γ phonons driven by circularly-polarized laser ($I_0$=1.2×10$^{10}$ W/cm$^2$, λ=1907.5nm) with negligible instrumentation broadening ($\sigma_E$=0.01 eV). (b) Same as in (a), but for $I_0$=4.3×10$^9$ W/cm$^2$, λ=2530nm. (c) Same as in (a), but with wider instrumentation broadening. (d) Same as in (c) but for the laser conditions in (b). Black lines denote equilibrium lattice Floquet bands (phonon-free). Dashed lines represent equilibrium system Dirac cone. The color code is normalized to the phononic snapshot weight (without ARPES matrix elements or electronic occupations).

From Fig. 1(a,b) several results emerge. The first is obvious, which is that phonons induce a natural broadening of Floquet bands. This is in-line with past phenomenological or simplified e-ph coupling theories [29,37]. Moreover, broadening occurs mostly around Floquet bands associated with the equilibrium lattice. The second novel observation is that phonon occupations do not simply cause trivial broadening. There is a wide and almost $k$-point independent formation of Floquet states below the minimal (above the maximal) energy of the Floquet conduction (valence) band. This causes an effective reduction of the gap size by factor ~2, though the intensity of these modes is weaker compared to the dominant spectral features. The origin of these states is analyzed below. Further, we note the appearance of sharp modes that form flat-band-like features within the Dirac cone (e.g. at energies ±0.25 eV in Fig. 1(a), broadly appearing in all examined conditions). These are not flat bands *per se* since they arise from an average of many electronic Floquet bands shifted by phonon distortions, rather than a single band. Still, they signify that optical phonons strongly couple to this energy and $k$-region [54]. These features appear even at zero temperature since the dominant phonon contribution in graphene are zero-point modes, meaning temperature reduction cannot counteract them.

Moving forward, we repeat this analysis with a finite instrumentation broadening. We take rather optimistic values, $\sigma_E$ = 0.04, 0.05 eV in both cases, corresponding to ARPES probe pulse durations of ~80 fs. These values are chosen in each case as values that should still lead to a clearly discernible gap, where broadening is about half of the equilibrium lattice Floquet gap. Figure 1(c,d) presents resulting spectrograms, with the picture now much worse: The Floquet gap effectively closes due to the further coupling to instrumentation broadening. We define a visibility parameter as the ratio between the minimum of the spectral function $f(\mathbf{k}, E)$ at the gap, and maxima of nearby Floquet conduction/valence bands, which in this case reduces to $v$~1.6 (as opposed to $v \to \infty$ in absence of e-ph coupling, or $v \sim 70$ in absence of additional instrumentation broadening). This striking effect occurs regardless of the broadening being smaller than the gap. Indeed, we choose $\sigma_E \approx \Delta_{eq}/2$ (with $\Delta_{eq}$ the equilibrium lattice Floquet gap), which from pure gaussian broadening should lead to $v$~5. However, $\sigma_E$ couples to the phonon modes that reduce the gap, greatly diminishing visibility. Such nonlinear coupling also substantially smears the Floquet conduction/valence bands (Fig. 1(a,c)), and essentially fully closes the gap opening in the first hybridization point away from K (at $\Delta k \sim 0.06$ a.u.).

To study the mechanism that leads to this blurring, we re-analyze the negligible instrumentation broadening case. Figure 2(a) presents a decomposition of the individual phononic snapshots that make up Fig. 1(a). Each blue line represents a single Floquet band from a given $\Delta\mathbf{R}$ that was drawn out of Bose-Einstein Γ-phonon statistics. As seen, the snapshots completely fill the energy diagram, leaving only small gaps at the original position of the Dirac cone (the factor two gap reduction seen in Fig. 1). Further analyzing this result, we note avoided crossings between Floquet states of different lattice distortions (highlighted by purple lines). These can be analytically obtained as the average quasi-energies between pairs of consecutive bands in the equilibrium lattice case, and can be understood as phonon-preserved avoided crossings since here the Floquet Hamiltonian only couples to phonons off-diagonally in the $t_1$ term. Similar physics arises along $k_y$ (see SI).

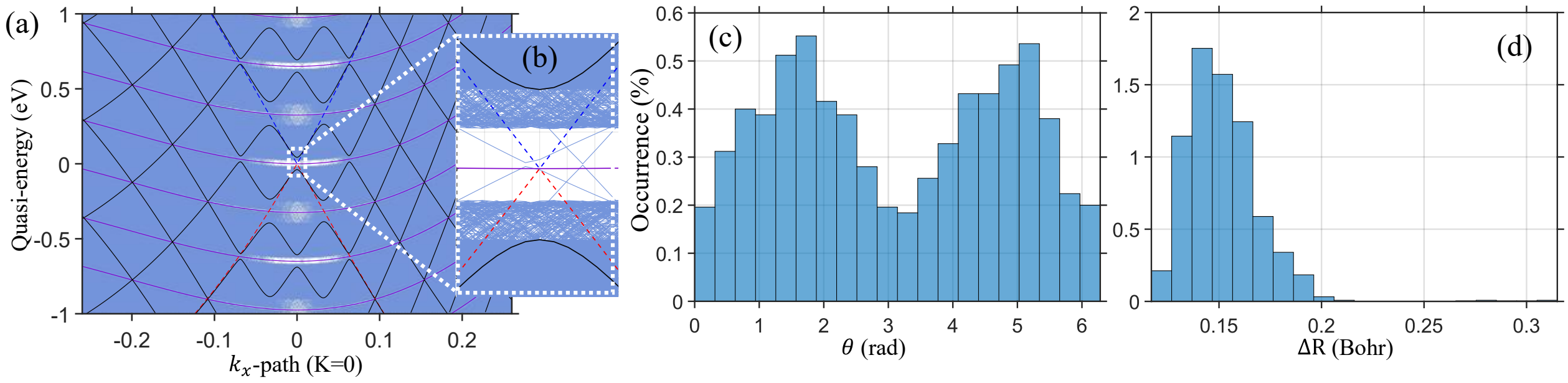

**Fig. 2.** Mechanism of Floquet gap blurring by e-ph coupling. (a) Snapshot-resolved Floquet bands (blue). Black curves indicate Floquet bands of the equilibrium lattice and dashed lines the original Dirac cone. Avoided crossing (AC) parabolic dispersion bands (purple) limit the quasi-energy domains of individual snapshot bands. (b) Zoom-in in gap region showing a reduced density of Floquet bands going below (above) the conduction (valence) Floquet band of the equilibrium lattice case, blurring the gap beyond simple broadening. (c,d) Statistics of snapshots reducing the gap, showing broad angular spread of displacements (with slight preference for transverse modes at $\theta=\pi/2, 3\pi/2$), and a minimal threshold of required lattice distortion with a main peak around 0.15 Bohr. Note that in (c) and (d) a larger snapshot statistics of 25,000 cases was used to obtain smoother histograms.

Connecting to the individual snapshots, we analyze the states responsible for the effective gap reduction. Figure 2(b) shows a zoom-in around the gap, indicating that the density of modes that manage to traverse the gap (i.e. 'enter the gap') is much smaller than the dominant modes that linger around the equilibrium lattice Floquet bands. We extracted the phonon displacements that gives rise to these bands and found that they reflect lattice distortions of the C-C bond larger than ~0.05 Å, and typically maximized at ~0.08 Å (see Fig. 2(d)). The directionality of the distortion is less significant and is broad, with slight preference for a transverse distortion (i.e. $y$-axis distortion is preferred along $k_x$ line gap closure, and vice versa, see Fig. 2(c)). The occurrence of these modes is roughly 10% of all optical Γ phonons, since they arise just beyond one standard deviation of the phonon statistics. These are the modes responsible for the reduction in Floquet gaps, and also are the ones that couple to instrumentation broadening for large $\sigma_E$ (see SI). Physically, they arise because for sufficiently large $\Delta\mathbf{R}$ the Floquet Hamiltonian structure changes, shifting the equilibrium Dirac cone position (though not by enough to merge K and K' cones [55–57]), allowing gap size reduction due to band hybridization at larger momenta. In the SI we directly show that removing these large displacement modes re-opens the gap and also reduces the flat-band-like feature.

We next explore various dressing conditions. Figure 3 shows Floquet spectral functions for select dressing field wavelengths and intensities. The results show largely similar effects, even in much larger laser intensity or wavelength (which opens a large Floquet gap). For example, in Fig. 3(d) we have chosen $I_0=10^{10}$ W/cm$^2$, λ=2530nm, leading to a large equilibrium lattice Floquet gap of $\Delta_{eq}$~0.17eV. Still, the spectral function with e-ph coupling reduces the gap to Δ~0.07 eV. Generally, all tested conditions give rise to similar features with an identical origin (see SI for additional spectra), except that the required lattice distortions for 'entering the gap' differ – larger distortions are needed for scenarios that originally have smaller equilibrium Floquet gaps, while for large equilibrium Floquet gaps even minor distortions reduce the gap. This readily arises from the structure of the Floquet Hamiltonian, and is behind similar reductions in gap sizes across conditions. Generally, in all cases a factor two reduction in gap occurs (Fig. 3 (e,f)). Note that when the original $\Delta_{eq}$ is relatively small (e.g. on a scale <50meV), even much smaller and idealized instrumentation broadening completely closes the gap due to this effect.

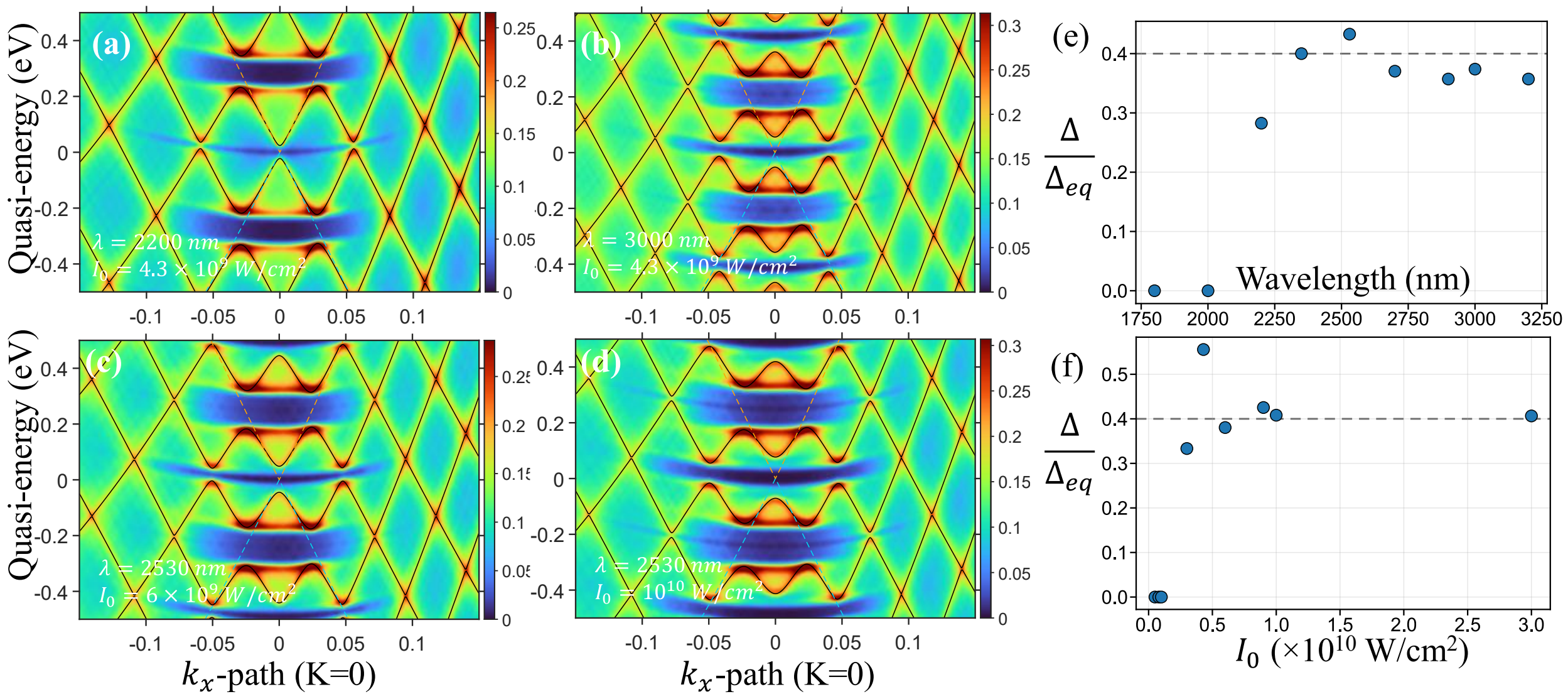

**Fig. 3.** Intensity and wavelength dependence of Γ-phonon induced Floquet gap blurring. (a)-(d) Select cases displaying substantial intra-gap Floquet mode presence ($\sigma_E$=0.01 eV). (e,f) Ratio of gap reduction, comparing practical gap Δ, to the equilibrium lattice Floquet gap $\Delta_{eq}$, across wavelengths (with a fixed intensity of 4.3×10$^9$ W/cm$^2$) and intensities (with a fixed wavelength of 2530 nm).

At this point, we recall that we limited eq. (2) to Γ-only phonons. This accounts for a major mean distortion in the graphene lattice. However, additional distortions caused by non-Γ modes are also relevant. In that respect, the results above can be seen as a lower limit of gap blurring, since e-ph blurring can only be enhanced by additional pathways (the summation of different channels is necessarily incoherent). Non-Γ modes can break inversion symmetry and tune the Fermi level position since they also modulate the $t_2$ term. These effects can give rise to qualitatively different features that might accelerate the above effect.

To test this, we extend our analysis to a 2×2 graphene supercell (yielding an 8×8 equilibrium Hamiltonian). This Hamiltonian is treated in the same manner, with the simplifying assumption that the phonon statistics are drawn out independently for each atom (see SI for details). Figure 4 shows a comparison of several cases with laser parameters as in Fig. 1(a): Fig. 4(a) presents results with Γ-only phonons at a single unit cell (from Fig. 1(a)), focusing on the line-cut at $K$ ($f(\mathbf{k}=K,E)$). The reduction in gap size is visible, though the gap is still 'closed' and the visibility is high ($v=71.8$). Figure 4(b) shows equivalent results from the supercell case, where phonons have been drawn from a restricted distribution of Γ-only modes. In this case slight additional blurring reduces the visibility to $v$ =16.6, which is a result of down-folded band replicas. This is taken as a benchmark for the non-Γ case to ascertain effects that are unrelated to band folding (which are difficult to remove and are essentially artifacts of the methodology). The key result from the supercell is shown in Fig. 4(c), including both Γ and non-Γ modes. The gap there is fully closed and the visibility parameter is reduced to ~7.2 (2.3 times lower than with only Γ phonons).

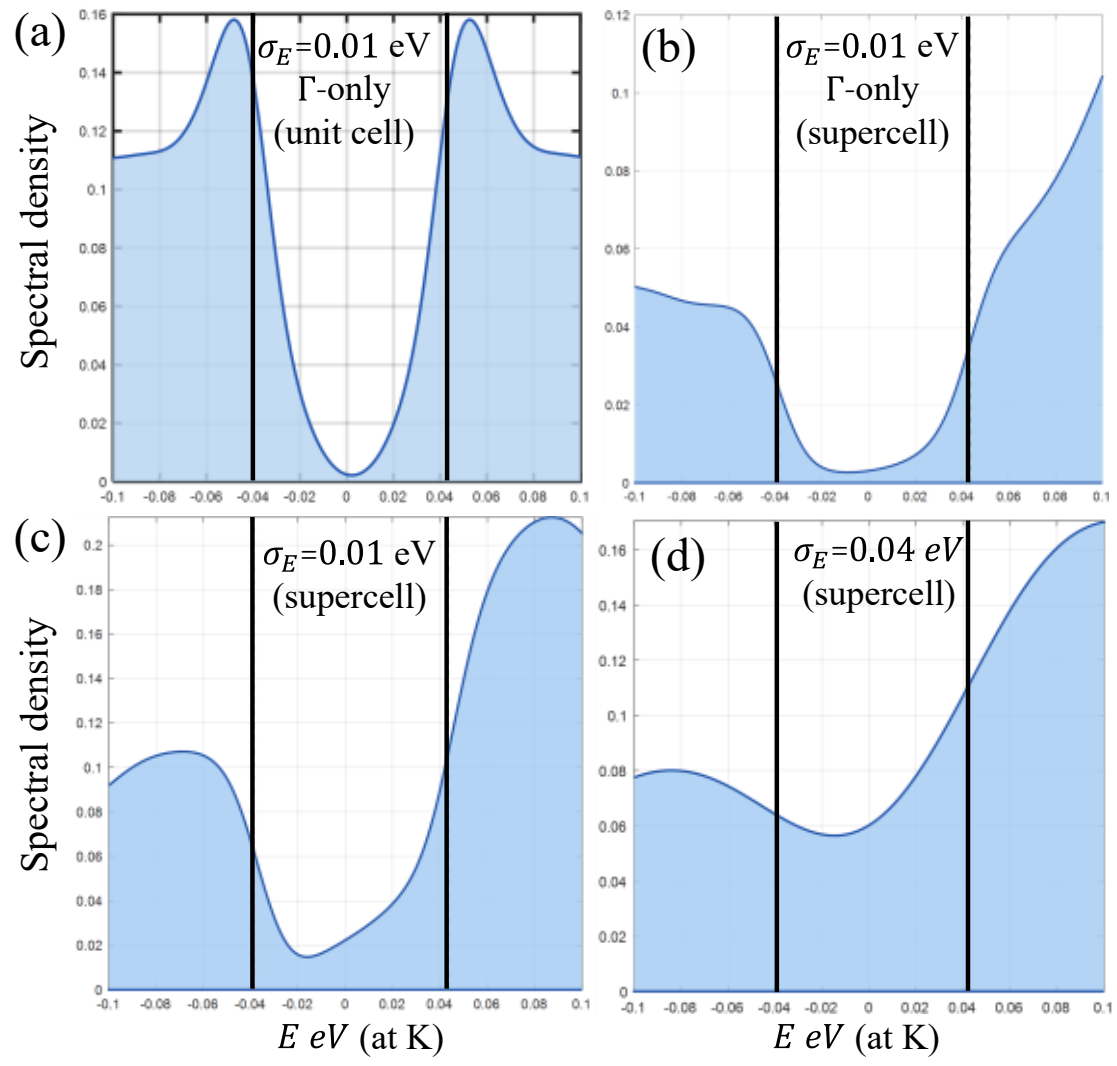

**Fig. 4.** Floquet topological gap blurring due to non-Γ phonons in similar conditions to Fig. 1(a). (a) Spectral function line-cut at K. Γ-modes reduce the gap size, but do not fully close it, leaving a still high visibility. (b) Same as (a) but for the supercell calculation where Γ-only modes are enforced. (b) is directly comparable to (a) except that some additional broadening arises from folded band replicas in the supercell. (c) Supercell case including non-Γ modes comparable to (b), showing reduced visibility. (d) Same as (c) but with instrumentation broadening that reduces visibility beyond observability. Vertical black lines denote Floquet band edges in the equilibrium case.

Physically, non-Γ phonons break inversion symmetry, allowing additional Floquet states to 'enter the gap'. In the SI we determine that this is a result of non-Γ modes coupling to the $t_1$ term rather than $t_2$, making apparent that it is not chemical potential shifts that causes the added blurring, rather, their broadened distribution and symmetry-breaking. Further coupling these modes to instrumentation broadening of 0.04 eV exasperates these features and completely blurs the gap, reducing visibility to 1.4 – a situation which is barely resolvable experimentally. In the SI we show that in these conditions visibility gradually reduces with $\sigma_E$ until at $\sigma_E$~0.04 eV the gap becomes undiscernible. Importantly, we emphasize that the analyzed spectral functions do not include effects of electron occupations or ARPES matrix elements, and it's unclear if such effects might further blur the gap or counteract the phononic features. However, due to the incoherent summation of all channels we expect those effects to further reduce gap visibility.

Overall, phonons cause Floquet gap visibility to reduce from essentially infinite in the equilibrium lattice non-broadened case, to ~70 including Γ phonons, to ~7 including also non-Γ modes, and to ~1.4 if that also incorporates reasonable probe broadening. Additional channels not considered (phononic, electronic, and experimental errors) likely reduce the number further. We note that this depends on the laser regime, and for instance more intense fields can lead to larger visibility.

To summarize, we theoretically explored Floquet topological physics in graphene driven by circularly-polarized lasers. Coupling electron dynamics non-phenomenologically to phonons, we showed that e-ph coupling drastically affects the spectral function (inducing flat-band-like features) and Floquet gap visibility. This arises via: (i) Large displacement Γ phonons that 'enter the gap' and reduce its size. (ii) Non-Γ modes that break crystal symmetries and close the gap and reduce its visibility. (iii) Coupling of the above two mechanisms to reasonably expected instrumentation broadening, which seriously exasperate the effect. Altogether, these results propose a reasonable explanation to the lack of Floquet topological gap observability in graphene. Perhaps just as importantly, our real-space analysis suggests that is not necessarily a limiting factor for electronics, since in local spatial crystal regions one has well-defined gapped modes and anomalous

Hall physics survives [22,24]. The effect on protected edge states is not as obvious, and while beyond our scope, should motivate future research.

Looking forward, our work naturally suggests possible mitigation strategies, such as pre-pumping coherent phonons that would phase-synchronize the lattice, following which a dressing field would induce a Floquet topological state, while a third ARPES probe would measure the bands. This scenario should remove non-phenomenological phononic effects, leaving only trivial broadening. Another solution could be to transition to a Dirac system with more favorable phonon statistics and less dominant zero-point modes, such as the recently predicted $B_2S$ [58], whereby in a low temperature limit phononic blurring might be mitigated. Lastly, simply employing higher peak powers would also be effective, though then the main concern is material damage.

**ACKNOWLEDGMENTS**

The authors thank Prof. Angel Rubio and Prof. Umberto De Giovannini for insightful discussions. O.N. gratefully acknowledges the Young Faculty Award from the National Quantum Science and Technology program of Israel's Council of Higher Education Planning and Budgeting Committee and the Technion NEVET programs of the RBNI and Helen Diller Quantum Center.

# Supplementary Material

## TECHNICAL DETAILS

We report here on technical details for the calculations presented in the main text. We begin with details of the simulations involving the single unit cell with Γ e-ph coupling. Spectrograms in the main text were computed on $k$-grids sampling a total 600 k-points along $k_x$ or $k_y$, passing the $K$ point in its center, with a total size of $b/3$, with $b$ the size of a reciprocal lattice vector. Converged simulations include 10,000 phonon snapshots unless stated otherwise (see convergence comparison below in Fig. S1).

The specific form of the 2×2 Hamiltonian included the following expression for the modified hopping amplitudes per the $i$'th NN bond: $\tilde{t}_{1,i}(\Delta\mathbf{R}) = t_1 \exp\left(-\beta\left(\frac{|\boldsymbol{\delta}_{i,1}+\Delta\mathbf{R}|}{|\boldsymbol{\delta}_{i,1}|} - 1\right)\right)$, where $\beta = 3$ is a typical value in graphene [59]. We then had the modified spectral function take the form $\tilde{f}_1(\mathbf{k}, \Delta\mathbf{R}) = \sum_i \tilde{t}_{1,i}(|\Delta\mathbf{R}|)\exp\left(-i\boldsymbol{\delta}_{i,1} \cdot \mathbf{k}\right)$, with the summation running over the three NN sites (note that the phase shift of the structure function has been ignored for simplicity since it does not change the Floquet eigenvalues). The $t_2$ term is unmodified with Γ-only phonon modes. The displacements $\Delta\mathbf{R}$ were drawn out following the procedure in ref. [53] using Bose-einstein statistics only for the LO and TO modes (neglecting out-of-plane modes). We used a graphene lattice parameter of 2.46 Å, and hopping values $t_1 = -2.4205$ eV and $t_1 = 0.4462$ eV.

For each snapshot (**ΔR**) we expanded the 2×2 ground state Hamiltonian in the Floquet extended space [49] with up to 16 photonic channels in a numerical approach equivalent to that in refs. [51,52]. The Floquet integrals were solved numerically with Simpson integration using 150 time points per optical cycle. Note that these parameters are well above typical convergence criteria and are only needed for very intense laser fields (in most laser settings no more than 4 photonic channels are needed). Each Floquet Hamiltonian was numerically diagonzlied to obtain Floquet quasi-energy bands, as in Fig. 2(a) in the main text.

To obtain the phononic Floquet spectral unction, $f(\mathbf{k}, E)$, each Floquet quasi-energy band in each snapshot was replaced by a spline interpolant 2D function along the $k$-path with a natural energy-broadening of with $\sigma_E$ of the form $f_{n,j}(\mathbf{k}, E) = \exp\left(-\frac{\left(E-E_j(\mathbf{k})\right)^2}{\sigma_E^2/4\ln 2}\right)$, with $E_j(\mathbf{k})$ a spline interpolant of the quasi-energy Floquet band dispersion of the $j$'th band in the $n$'th snapshot. The full spectral function was obtained by summing over all Floquet bands indices and snapshot averaging: $f(\mathbf{k}, E) = \frac{1}{N_{ch}N_{snap}}\sum_n \sum_j f_{n,j}(\mathbf{k}, E)$, with $N_{snap} = 10{,}000$ the snapshot number, and $N_{ch}$ the number of photonic channels used.

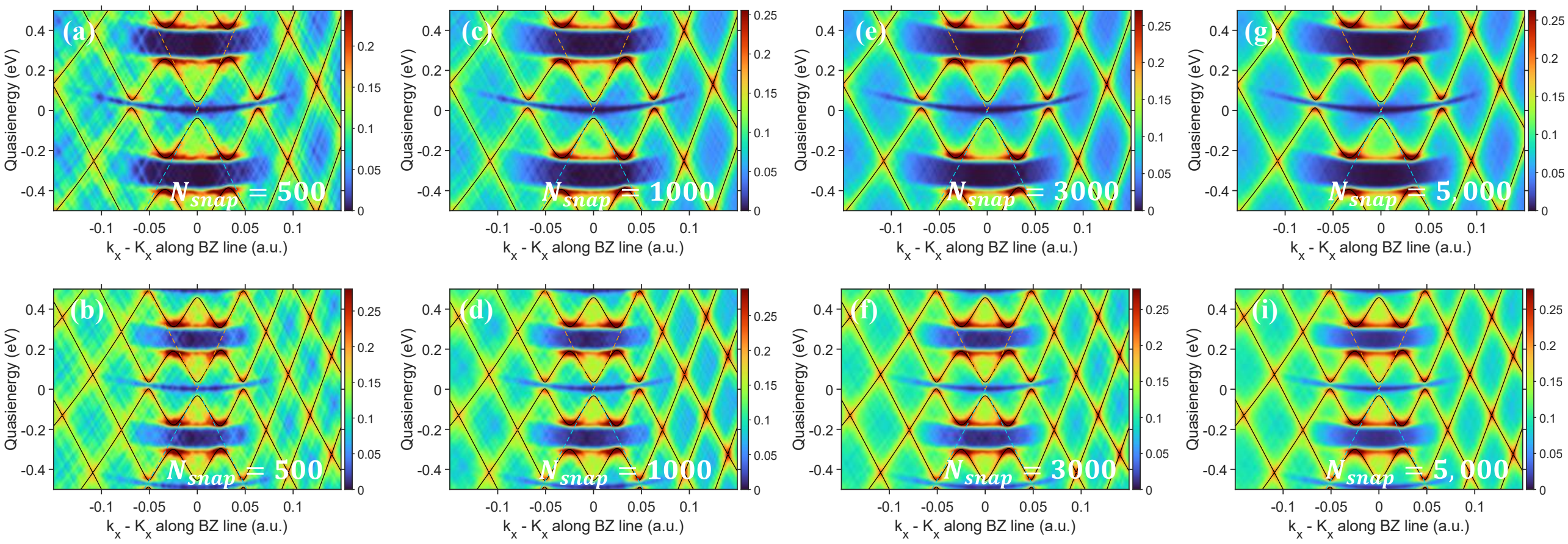

FIG. S1. Convergence testing in the single unit cell case vs snapshot number in similar conditions to Fig. 1(a) in the main text (top 1,c,e,g) and Fig. 1(b) in the main text (bottom, b,d,f,i).

In the 2x2 supercell case we employed a similar strategy. The supercell Hamiltonian had an 8×8 form that was calculated up to 2nd NN terms with similar values. In equilibrium and absence of laser driving this produces the folded band structure seen in Fig. S2(a,c), with a clear Dirac cone dispersion at $K$. The NN hopping was modified in an identical setting, except that the NN hopping vectors were defined for each of the separate 8 lattice sites in the supercell, modifying the resulting Hamiltonian terms. The 2nd NN term was modified in a similar manner via the function: $h(x) = t_2 \exp(-\alpha x^2)(1 + a_2 x^2 + a_4 x^4 + a_6 x^6 + a_8 x^8 + a_{10} x^{10})$, where the hopping takes the normalized form for the $j$'th 2nd NN connection: $\tilde{t}_{2,j}(\Delta\mathbf{R}) = t_2 \frac{h(|\boldsymbol{\delta}_{2,j}+\Delta\mathbf{R}|)}{h(|\boldsymbol{\delta}_{2,j}|)}$. The parameters $a_{2m}$ were taken as $a_2$=8.77×10$^{-3}$, $a_4$=1.53×10$^{-4}$, $a_6$=9.74×10$^{-7}$, $a_8$=−7.034×10$^{-9}$, $a_{10}$=8.813×10$^{-11}$, and we used $\alpha$=3.33×10$^{-2}$. These parameters were taken from a separate calculation of a hydrogenic orbital $p_z$-$p_z$ orbital semi-analytic overlap integral with increasing distance in the nodal plane, as occurring on a honeycomb lattice. This integral mimics the expected decay behavior of the hopping elements with distance in absence of more accurate data as we have for the $t_1$ term. The exact numeric result was fitted with a least-squares approach to the above form of a decaying gaussian multiplied by an even order polynomial, yielding essentially an exact match. In the supercell simulations the $f_2$ structure factor was modified based on the attenuation of the 2nd NN bonds, just as was done for $f_1$ in the single unit cell case. From this Hamiltonian, we expanded the Floquet Hamiltonian in a similar manner to the single unit cell. In this case, since we limited our simulations to the laser regime in Fig. 1(a) we used less strict numerical parameters with maximal 10 photonic channels and 400 $k$-points along a similar $k$-path.

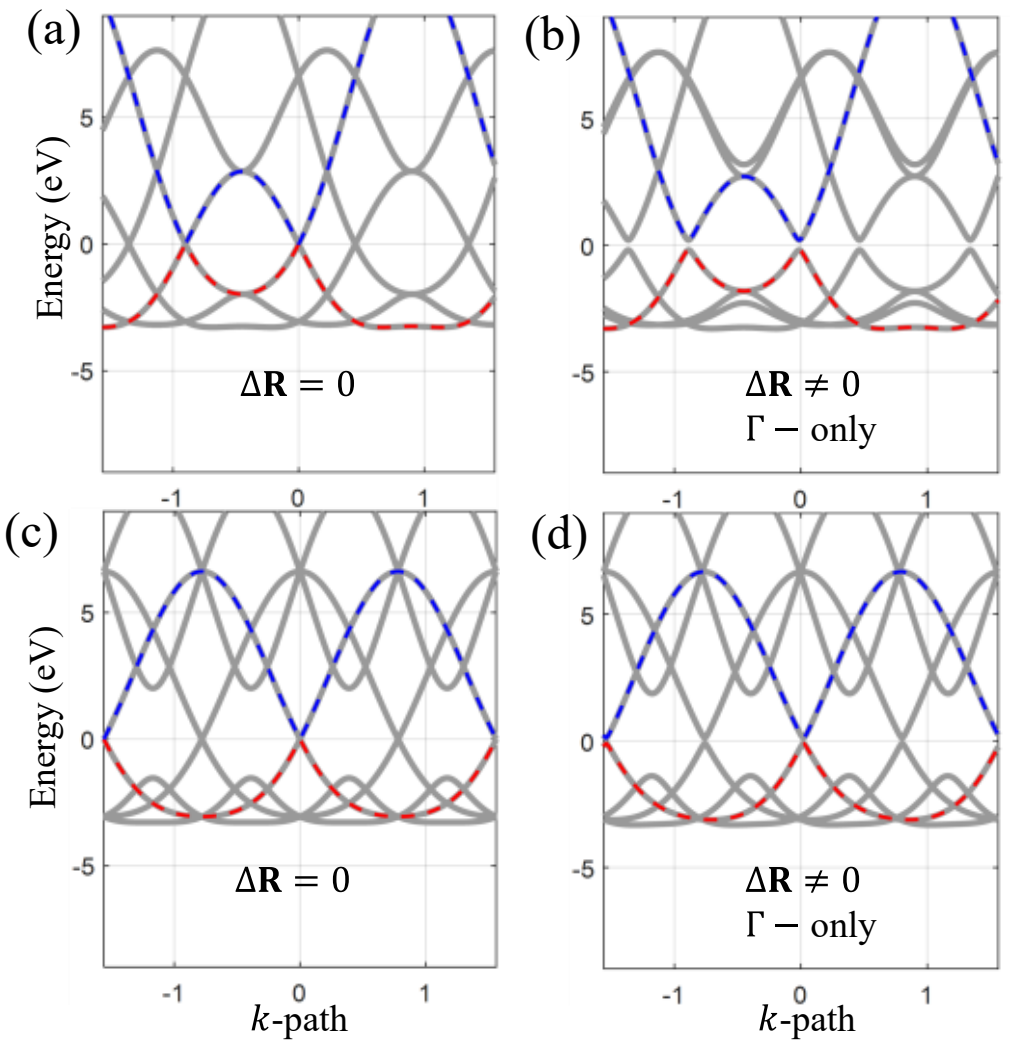

FIG. S2. 2x2 supercell band structure benchmark in absence of laser drive showing band folding. (a) $k_x$ dispersion bands in the 2x2 supercell BZ (grey), with the original overlaid 1x1 unit cell bands in dashed lines, for the perfect equilibrium lattice case, showing an exact mapping as expected. (b) Same as (a) but with a nonzero lattice displacement (exemplary phonon snapshot) that has been drawn from a Γ-phonon in both the unit cell (dashed) and supercell (grey), showing identical bands to the supercell-folded bands. (c,d) Same as (a,b) but along $k_y$ dispersion.

The lattice distortions were drawn out in a similar manner to the unit-cell case. That is, for each of the 8 carbon atoms in the supercell sitting at an equilibrium position $\mathbf{r}_m$ with $m$ the atomic index running from 1 to 8, we drew from a normal distribution lattice distortions $\Delta\mathbf{R}_m$ for each site in LO and TO modes. The statistics are taken from the Γ optical mode statistics as a simplification, while assuming each site has a harmonic and independent distribution from the other sites. This is equivalent to an Einstein approximation. In a realistic scenario these distortions are not independent and arise from the energies and momenta of the phononic band structure. We note however that in the 2x2 supercell we only access very minimal non-Γ phonons, and the simplification essentially assumes that these modes have similar energies for the statistical sampling. This approach in any case yields correct qualitative physics that gives rise to phonon modes that break inversion symmetries, mirror symmetries, etc. All supercell simulations used 15,000 phonon snapshots to reach convergence due to the added degrees of freedom requiring sampling. Fig. S2(b,d) shows examples of the non-laser-driven band structure for a randomly drawn out Γ phonon distorted lattice in the supercell case that is directly comparable to the unfolded unit cell case (both cases can be easily compared here due to the Γ nature of the distortion), which validates the Hamiltonian employed and acts as a sanity check for the approach.

## ADDITIONAL RESULTS

We present here complementary results to those in the main text. First, we present for the supercell case the phononic Floquet spectra where the $t_2$ term has been fixed despite allowing non-Γ phonons in the distribution (i.e. we have taken $\tilde{t}_{2,j}(\Delta\mathbf{R}) = t_2$). Figure S3 compares the frozen $t_2$ case with the typical one in Fig. 4(c) in the main text. Both results show essentially identical spectra, which pinpoint the non-Γ phonon effects causing gap closure not due to oscillations of the Fermi level (which is modified if $t_2$ varies since it lies on the 2×2 sub-block Hamiltonian diagonal), but rather by coupling to the $t_1$ more dominant term. In terms of energy scales this is completely intuitive – the $t_1$ term follows much larger variations with $\Delta\mathbf{R}$ due to its larger equilibrium value, and also due to the equilibrium NN bonds being much closer and therefore at a higher curvature region in the exponentially decaying function. Indeed, the $t_2$ term varies on a scale of ~0.01 eV for reasonable supercell phonon distortions, which is a minute effect.

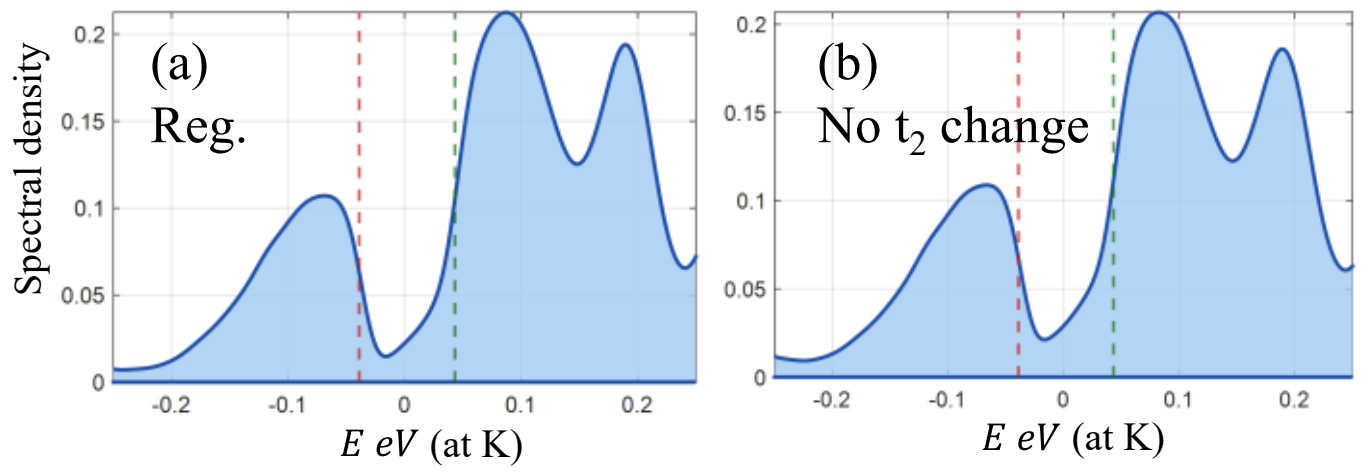

FIG. S3. Line-cut of spectral function for $k$=K in supercell for: (a) standard case as in Fig. 4(c) in main text, and (b) in similar conditions but where $t_2$ term is not allowed to modify, which fixes the chemical potential at 0 energy.

Figure S4 shows complementary line cuts at K along the energy axis for the spectral functions in the 2x2 supercell vs instrumentation broadening, showing how the visibility of the gap is gradually diminished, until the topological gap becomes undiscernible around $\sigma_E \sim 0.04\ eV$.

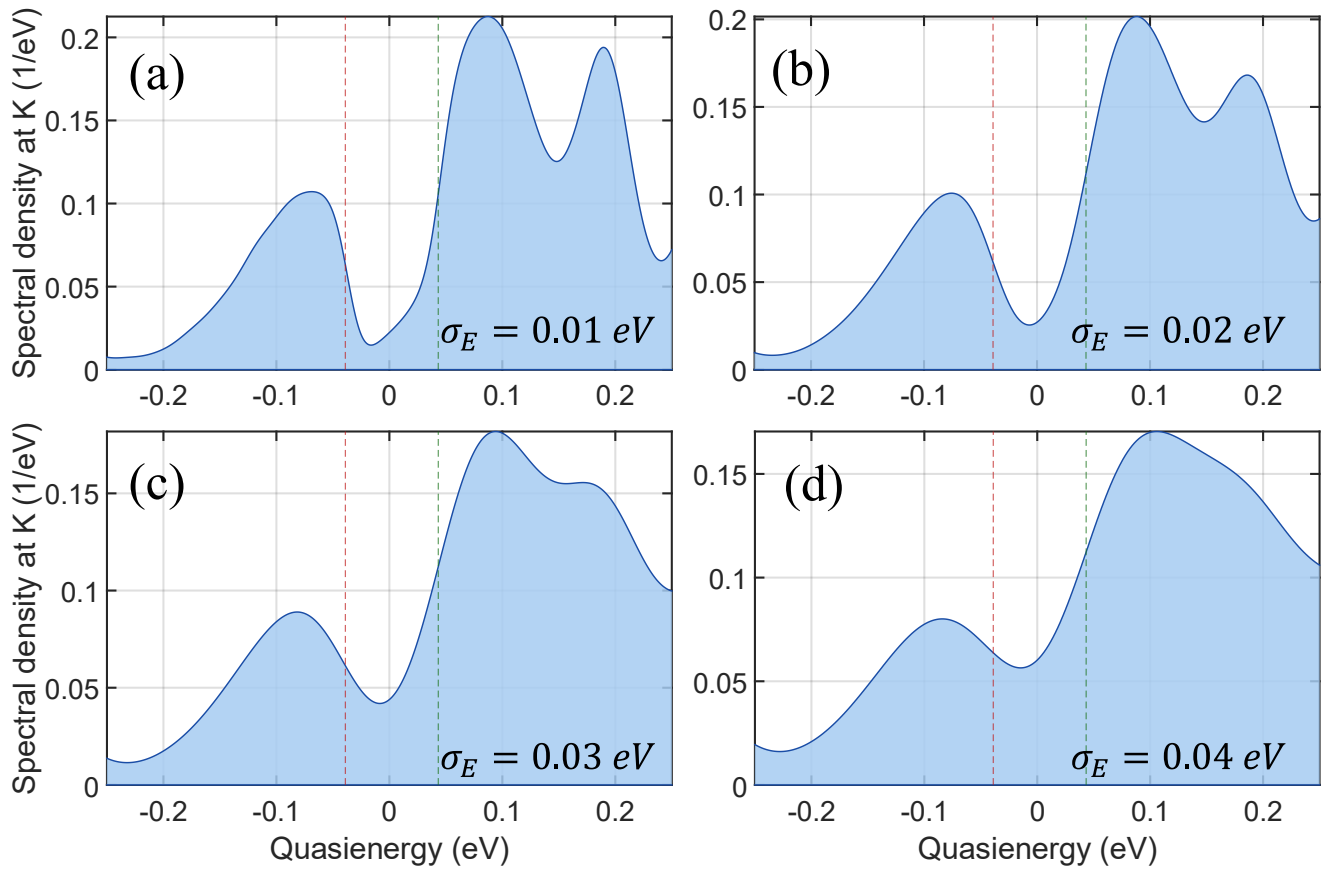

FIG. S4. Line-cut of spectral function for $k$=K in the supercell for (a) standard case as in Fig. 4(c) in main text, and (b-d) with gradually more substantial instrumentation broadening.

The full spectral functions in the supercell case corresponding to the cases analyzed in the main text (Fig. 4(b,c)) are presented in Fig. S5. The data show additional blurring induced by folded Floquet bands that is very difficult to unfold due to the entangled nature of the multi photonic channel system. Still, the Floquet gap is clearly discernible at an equilibrium lattice (as in Fig. S2), and Fig. S5 shows that Γ modes blur it slightly (Fig. S5(b), which is very much exasperated in the supercell case with non-Γ modes added as well (Fig. S5(a)).

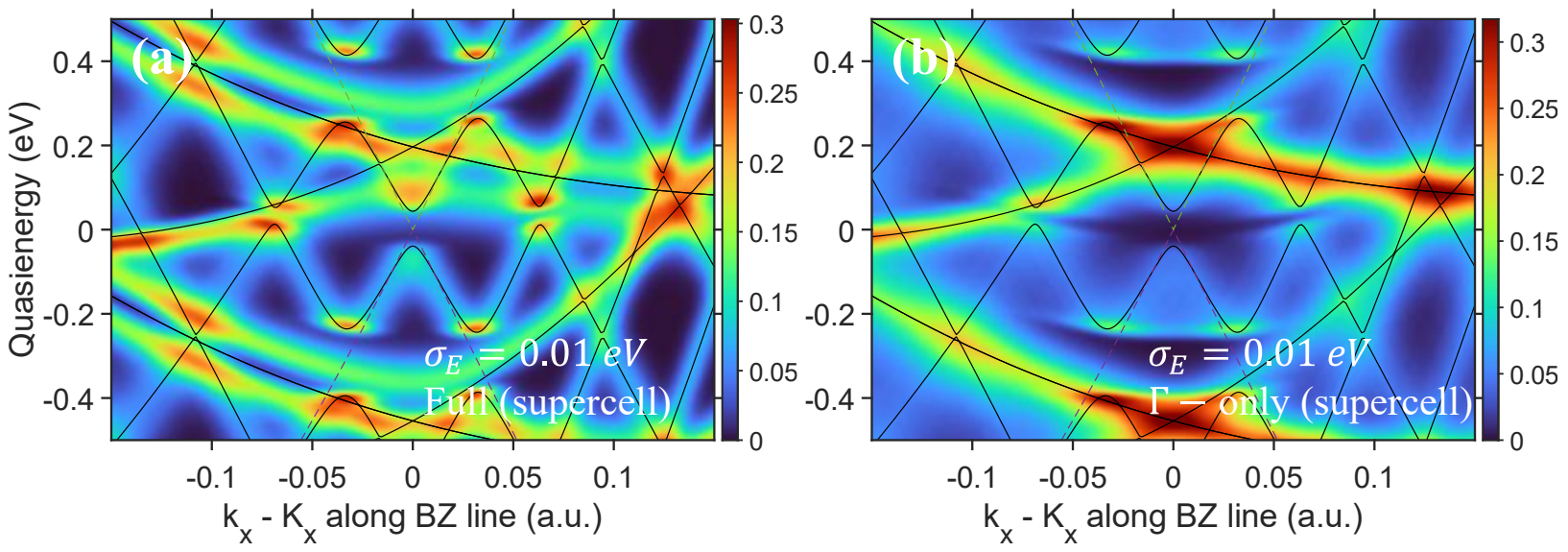

FIG. S5. Line-cut of spectral function for $k$=K in the supercell for (a) standard case like in Fig. 4(c) in main text, and (b-d) with gradually more substantial instrumentation broadening.

Next, for completeness, we present complementary $k_y$ Floquet spectra for Fig. 1 in the main text. This result is shown in Fig. S6. Identical physical effects and features are identified.

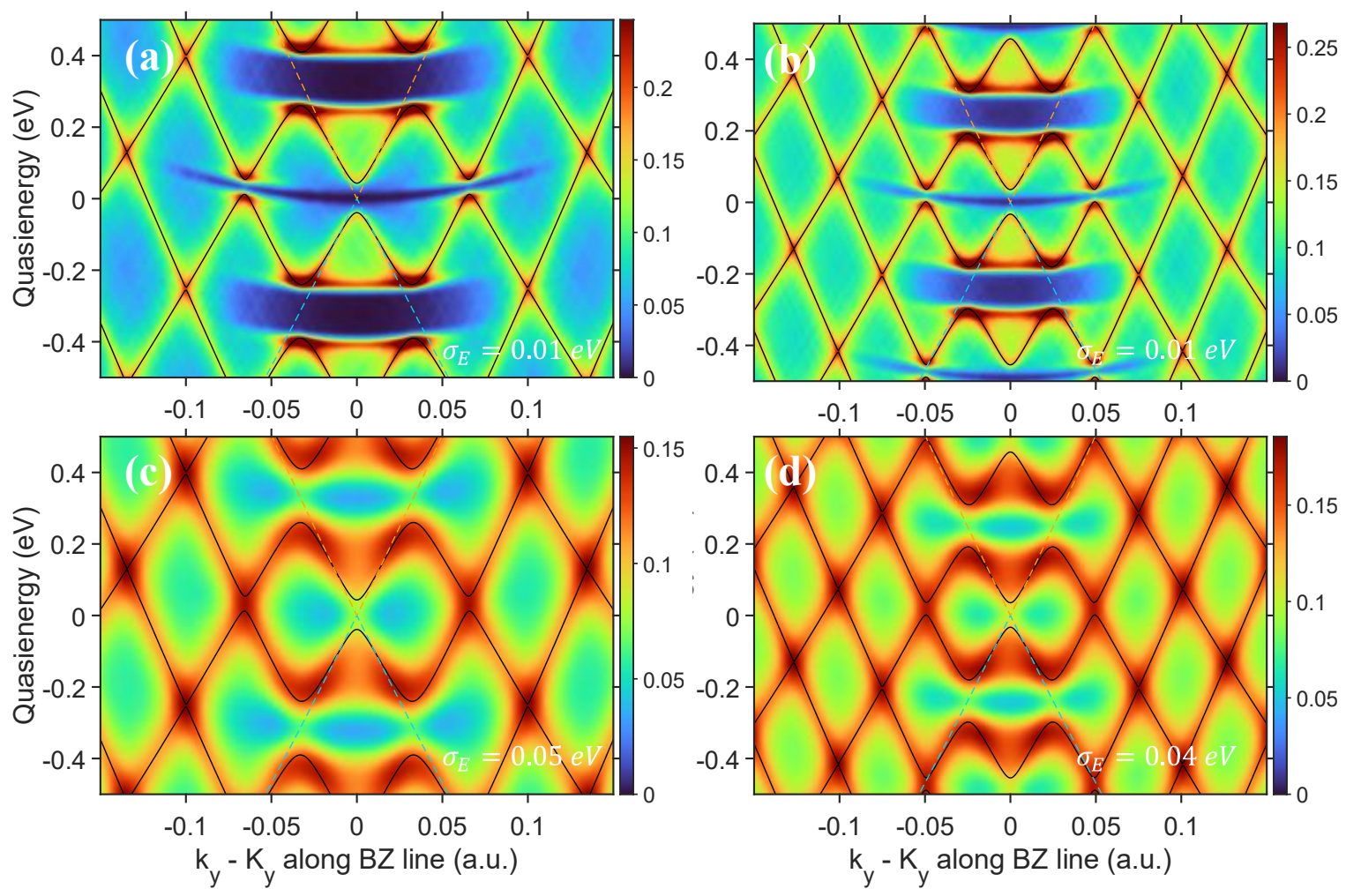

FIG. S6. Identical figure to Fig. 1 in main text, but along $k_y$ dispersion.

For completeness we also present all spectral functions for various laser wavelength and intensity scans that relate to Fig. S3 in the main text, from which the reduction of gap sizes was extracted. This data is presented for the $k_x$ case only ($k_y$ spectra are extremely similar and not presented) in Figs. S5-S6.

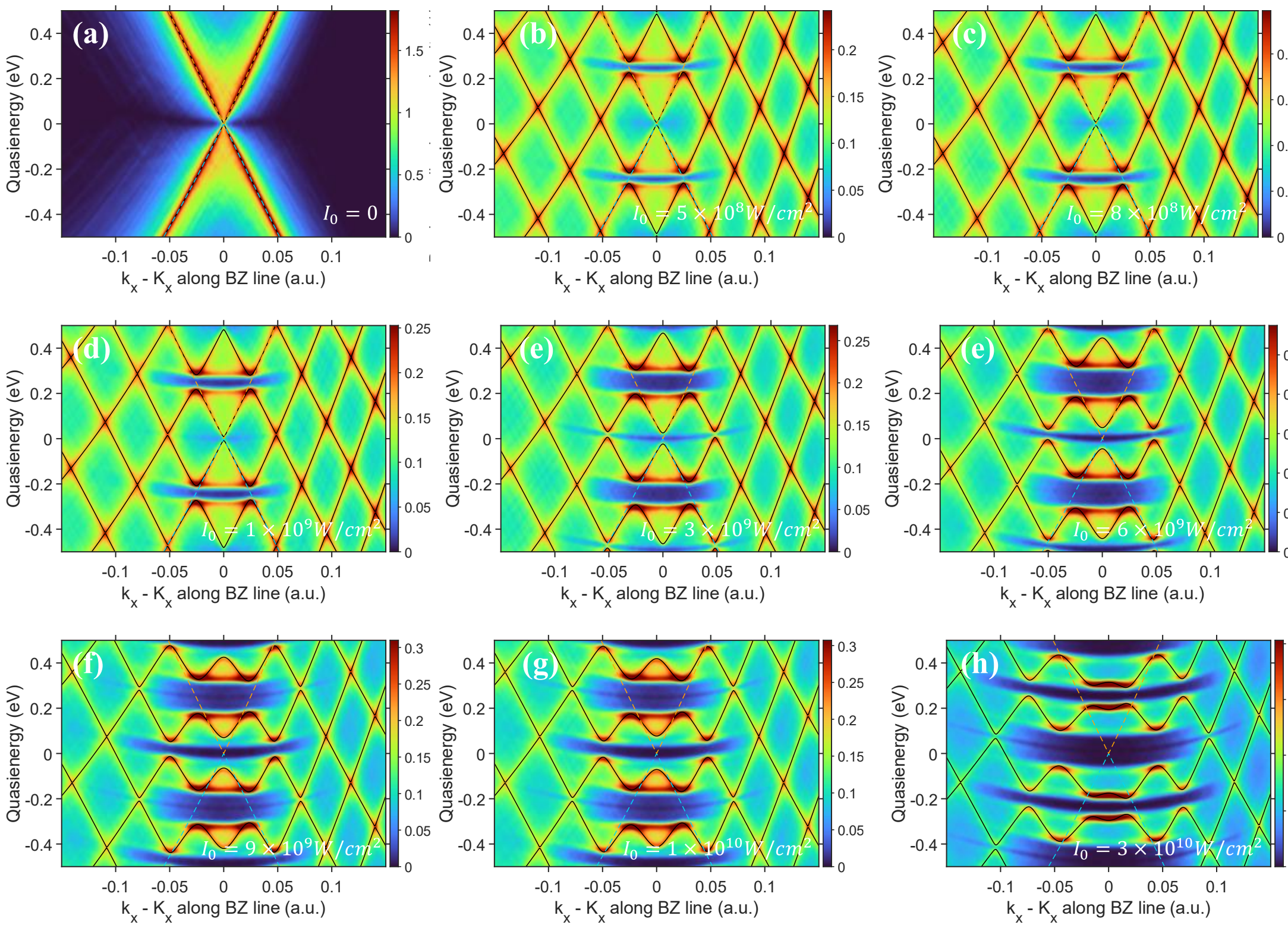

FIG. S5. Spectral functions for various intensities for a fixed wavelength of 2530nm, from which main text Fig. 3(f) was derived.

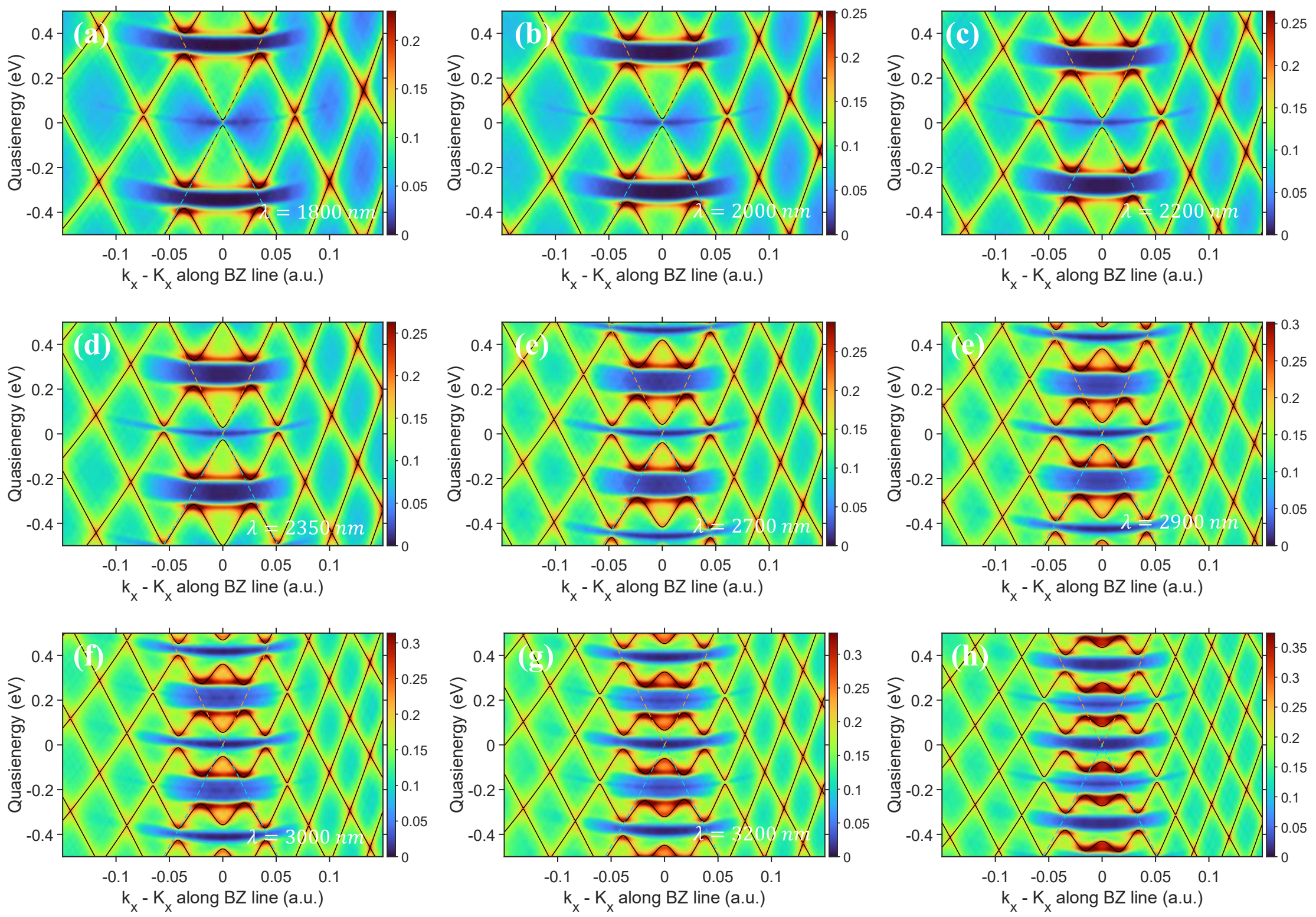

FIG. S6. Spectral functions for various wavelength for a fixed intensity of $4.3\times10^9$ W/cm$^2$, from which main text Fig. 3(e) was derived.

Lastly, we present spectra corresponding to Fig. 1(a,b) in the main text, but where the maximal phonon displacement was capped at roughly the phononic distribution FWHM (i.e. ~1 standard deviation). This result is shown in Fig. S7, and demonstrates that by removing large-displacement phonons there are no more Floquet bands entering the gap, as well as modes that form the flat-band-like feature.

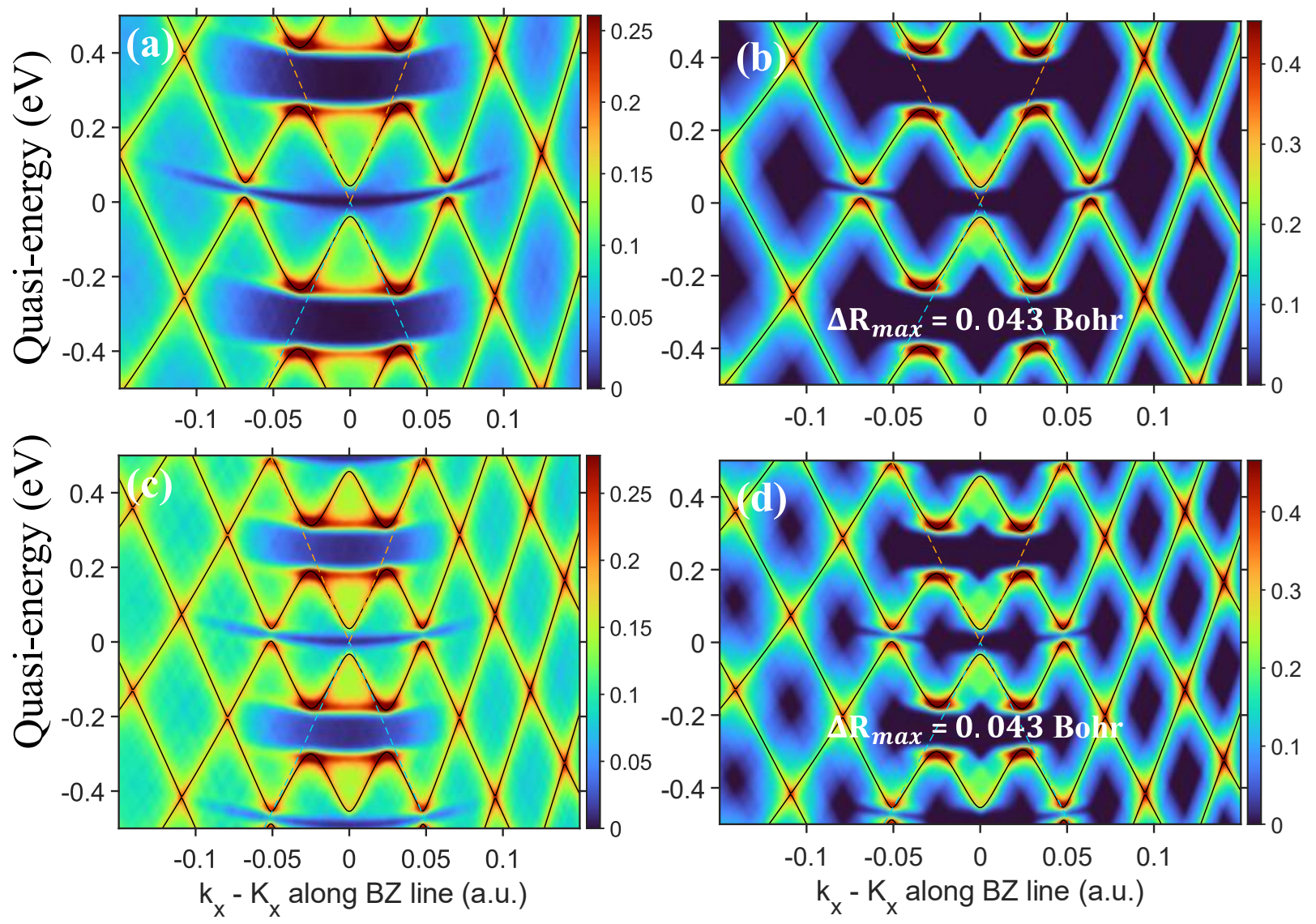

FIG. S7. Spectral functions in similar conditions to Fig. 1 in the main text. (a) Similar to Fig. 1(a) in main text. (b) Same as (a), but where the snapshots of phonon occupations were capped at one standard deviation, which removes the modes that 'enter the gap', as well as the flat-band-like feature discussed in main text and seen in (a). (c,d) Same as (a,b), but in conditions as Fig. 1(b) in the main text.